\documentclass[twocolumn]{aastex63}

\usepackage{etoolbox}
\usepackage{graphicx}	
\usepackage{multirow}
\usepackage{cleveref}
\usepackage{CJK}

\newcommand{\kms}{km s$^{-1}$}

\newcommand{\lam}{$\lambda$}

\newcommand{\lya}{\mbox{Ly$\alpha$}}

\newcommand{\civ}{\mbox{C\,{\sc iv}}}

\newcommand{\cii}{\mbox{C\,{\sc ii}}}
\newcommand{\siiv}{\mbox{Si\,{\sc iv}}}

\newcommand{\siii}{\mbox{Si\,{\sc ii}}}
\newcommand{\nv}{\mbox{N\,{\sc v}}}

\newcommand{\ovi}{\mbox{O\,{\sc vi}}}

\newcommand{\oi}{\mbox{O\,{\sc i}}}

\newcommand{\mgi}{\mbox{Mg\,{\sc i}}}
\newcommand{\mgii}{\mbox{Mg\,{\sc ii}}}

\received{}
\revised{}
\accepted{}
\submitjournal{ApJ}

\shorttitle{Absorption-Line Environments of High-Redshift BOSS Quasars}
\shortauthors{Chen et al.}
\graphicspath{{./}{figures/}}

\begin{document}

\title{Absorption-Line Environments of High-Redshift BOSS Quasars}

\correspondingauthor{Chen Chen, Bo Ma}
\email{chenchen8@mail.sysu.edu.cn, mabo8@mail.sysu.edu.cn}

\author{Chen Chen}
\affiliation{School of Physics $\&$ Astronomy\\
Sun Yat-Sen University\\ 
Zhuhai 519000, China}
\affiliation{Department of Physics $\&$ Astronomy\\
University of California\\ 
Riverside, CA 92521, USA}

\author{Fred Hamann}
\affiliation{Department of Physics $\&$ Astronomy\\
University of California\\ 
Riverside, CA 92521, USA}

\author{Bo Ma}
\affiliation{School of Physics $\&$ Astronomy\\
Sun Yat-Sen University\\ 
Zhuhai 519000, China}

\author{Britt Lundgren}
\affiliation{Department of Astronomy,\\
University of Wisconsin\\
Madison, WI 53706, USA}

\author{Donald York}
\affiliation{Department of Astronomy $\&$ Astrophysics,\\
University of Chicago\\
Chicago, IL 60637, USA}

\author{Daniel Nestor}
\affiliation{Department of Physics $\&$ Astronomy,\\
University of California\\
Los Angeles, CA 90095, USA}

\author{Yusra AlSayyad}
\affiliation{Department of Astrophysical Sciences,\\
Princeton University\\
Princeton, NJ 08544, USA}
\nocollaboration{7}



\begin{abstract}


The early stage of massive galaxy evolution often involves outflows driven by a starburst or a central quasar plus cold mode accretion (infall) which adds to the mass build-up in the galaxies. To study the nature of these infall and outflows in the quasar environments, we have examined the correlation of narrow absorption lines (NALs) at positive and negative velocity shifts to other quasar properties, such as their broad absorption line (BAL) outflows and radio-loudness, using spectral data from SDSS-BOSS DR12. Our results show that the incidence of associated absorption lines (AALs) and outflow AALs is strongly correlated with BALs, which indicates most AALs form in quasar-driven outflows. Multiple AALs are also strongly correlated with BALs, demonstrating quasar outflows tend to be highly structured and can create multiple gas components with different velocity shifts along our line of sight. Infall AALs appear less often in quasars with BALs than quasars without BALs. This suggests that BAL outflows act on large scale in host galaxies and inhibit the infall of gas from the IGM, supporting theoretical models in which quasar outflow plays an important role in the feedback to host galaxies. Despite having larger distances, infall AALs are more highly ionized than outflow AALs, which can be attributed to the lower densities in the infall absorbers. 


\end{abstract}



\section{Introduction}

High-redshift quasars identify episodes of rapid accretion onto super-massive black holes (SMBHs) in the center of massive galaxies. Possibly triggered by a recent merger of gas-rich galaxy, this rapid accretion activity is believed to be accompanied by rapid star formation during the early stages of galaxy evolution \citep{Sanders88, Elvis06, Hopkins08, Veilleux09}. Powerful outflows during this evolution stage, driven by the quasar and/or the starburst, expel gas and dust from the host galaxies, can quench the star formation, and cut off the fuel supply to the central black hole \citep{Silk98, Kauffmann00, King03, Scannapieco04, DiMatteo05, Hopkins08, Ostriker10, Debuhr12, Rupke13, Rupke17, Cicone14, Weinberger17}. The mechanism for the high-speed quasar-driven outflows quenching star formation in their host galaxies is expected to involve shredding and dispersal of interstellar clouds, which can produce complex outflowing gas structures on large scales \citep{Hopkins10, Faucher12b}. Infalling gas from the intergalactic medium (IGM) (e.g. cold mode accretion) is also believed to be important for fueling the central SMBH and star formation during these early active evolution stages \citep{Katz03, Keres09, Keres12}. 

Absorption lines in quasar spectra are unique tools to study the gaseous environments of quasars and test models of massive galaxy evolution. High-speed quasar-driven outflows are readily detected in quasar spectra via blueshifted broad absorption lines (BALs) with velocity widths larger than 2000 \kms\ \citep{Anderson87,Weymann91}. Low-speed outflows or infall in the extended host galaxies should produce narrow associated absorption lines (AALs) with redshifts near the quasar emission-line redshifts ($z_{abs}\approx z_{em}$) and velocity widths less than a few hundred \kms, \citep[e.g.,][]{Weymann79, Foltz86, Hamann97d, Hamann04, Simon10, Muzahid13}. Gas fragments shredded by powerful quasar outflows might produce rich multiple AALs \citep{Hopkins10, Faucher12b, Chen18, Chen19}. 


These different absorption-line environments should produce different signatures in the line kinematics, ionizations, column densities, and metal abundances. For example, infall lines at $z_{abs}> z_{em}$ generally have low metallicities and low ionizations since the gas coming from the IGM resides at large distances from the quasars. High-speed quasar-driven outflows originating in the galactic nuclei should exhibit the opposite behavior, with higher metallicities, higher ionizations, broader profiles, and stronger lines resulting from higher column densities. However, the complex nature of quasar/host galaxy environments could cause individual absorption-line systems to deviate from the typical behaviors that signal a particular origin or formation region. At negative velocity shifts, $z_{abs}< z_{em}$, the absorption lines that form in outflows are also mixed with unrelated lines that form in cosmologically intervening gas and galaxies along our line of sight. Thus it is hard to distinguish the origins of individual absorption lines. And when investigating the typical properties of quasars gaseous environments, it is better to use a large sample of absorption lines instead of individual lines. 



In this paper, we investigate the nature and origin of the diverse \civ\ narrow absorption lines (NALs) in high-redshift quasars measured in the Baryon Oscillation Spectroscopic Survey \citep[BOSS,][]{Dawson13, Paris17}, which is a part of the Sloan Digital Sky Survey III \citep[SDSS-III,][]{Eisenstein11}. This study is made possible by a new catalog of quasar absorption lines developed by York et al. (in prep.) using spectra from BOSS data release 12 (DR12). Our main focus is on the study of correlation between the incidence of \civ\ \lam 1548, 1551 NALs and other quasar properties.



The structure of this paper is organized as follows. Section 2 describes the quasar samples and the different types of NAL groups that we use in this study. Section 3 presents the main results based on both of the correlation analysis and the study of median composite spectra constructed for the different NAL groups. Section 4 discusses the results and the implications for our understanding of outflows and infall in the quasar environments. A brief summary is given in Section 5.

\section{Quasar Samples \& NAL Groups}

We select quasars based on the Quasar Absorption-Line Catalog from York et al. (in prep.), which is created using quasar spectra from BOSS DR12 \citep{Paris17}. The BOSS spectra have a wavelength coverage from $\sim3600$\AA\ through $\sim10,000$~\AA\ at a resolutions of $\sim1300$ in the blue end to $\sim2600$ in the red end \citep{Dawson13}. The selection criteria are that (1) the quasars must have redshifts in the range of $2.0<z_e<3.4$ to ensure the BOSS spectral coverage of \civ\ NALs at a wide range of radial velocity shifts (relative to the redshift of the quasar), and (2) median signal-to-noise ratio near the rest wavelengths of 1700~\AA\ (SNR$_{1700}$) is $\geq 3$. We have also rejected quasars with known damped \lya\ absorption lines (DLAs) in their spectra using the catalog flag of DLA\_flag$\,=-1$, which in the end yields a full sample of 100,376 quasars. 

From this full sample, we select well-measured \civ\ NALs for our study based on full width at half maximum (FWHM) $\leq$500 \kms, rest equivalent width (REW) $\geq$0.5~\AA\ measured at $\geq$3$\sigma$ significance, and \civ\ class score $\geq0.7$ recorded in the catalog. Both of the FWHM and REW cutoffs apply to the stronger \lam 1548 component in the \civ\ doublet. The \civ\ class score represents the quality assessments of the NALs identified from the machine learning algorithm in York et al. (in prep.). Based on these restrictions, more than 90\%\ of the \civ\ NALs included in our sample are reliable (see York et al., in prep., for more discussion).
The total number of quasars selected this way is 40, 696 from a previous full sample of 100,376. A total of 54,154 \civ\ NALs are well-measured in the spectra of this subsample. A cutoff value of REW at 0.3 or 0.4~\AA\ will yield a larger \civ\ NALs sample, but not alter the main results of the paper.

\begin{figure*}
\centering
\includegraphics[width=0.9\textwidth]{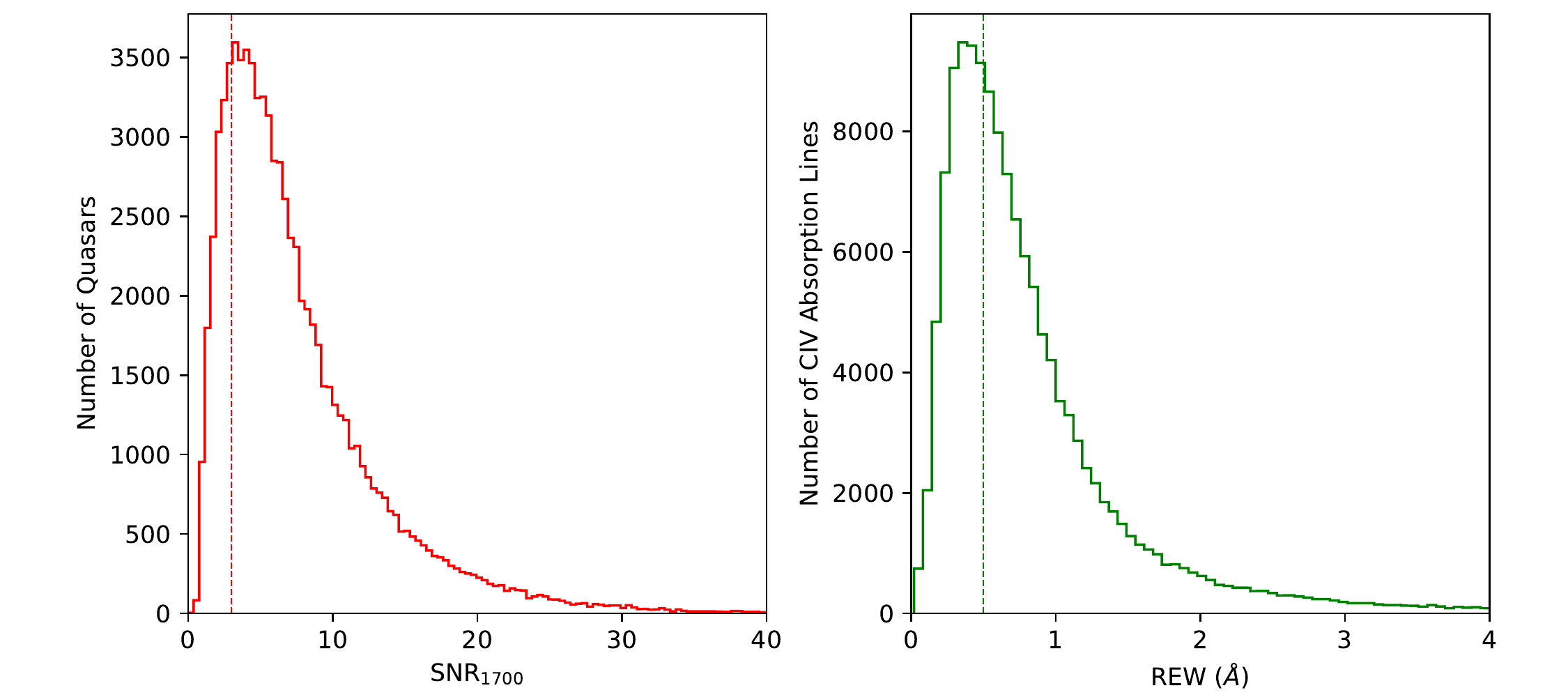}
\caption{Left panel: Distribution of SNR$_{1700}$ from spectra in our full quasar sample; Right panel: Distribution of REW from all the \civ\ absorption lines in our full quasar sample. The red and green dashed lines mark the thresholds at SNR$_{1700}=3$ and REW$\,=0.5$ \AA, respectively, which are used to select the subsample for this study.\label{fig:snr_REW_dis}}
\end{figure*}
\begin{figure}
\centering
\includegraphics[width=0.5\textwidth]{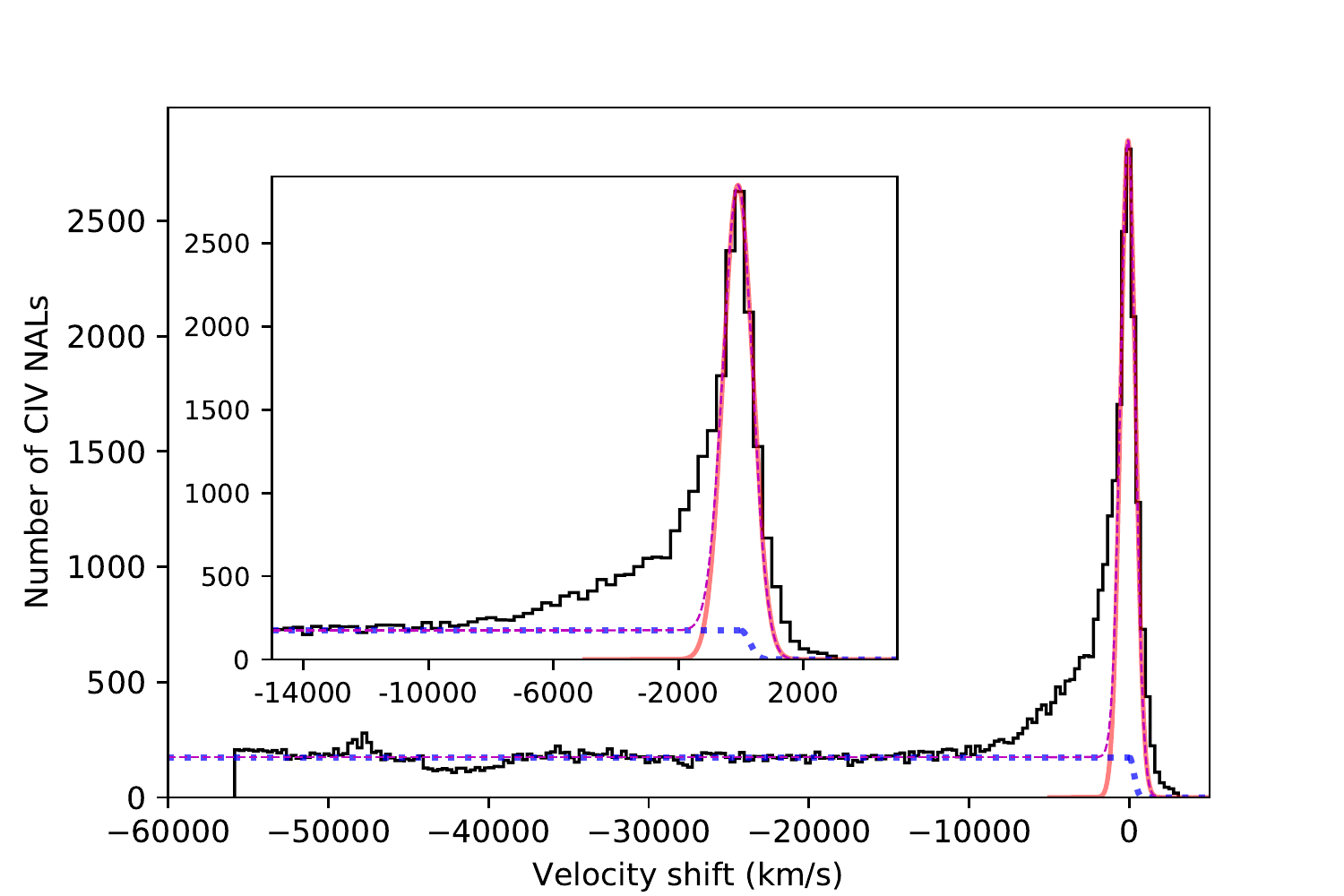}
\caption{Distribution of the velocity shifts in \civ\ NALs from our quasar subsample. The inset shows the expanded region of velocity range -15,000 to 5000 \kms, to show the details of the velocity distribution around zero velocity. The red solid curve and blue dotted curve are fits to the data in the velocity intervals $-1000< v < 1000$ \kms\ and $-60,000 < v < -15,000$ \kms, representing the random Gaussian distribution ($\rm{mean}=-93$ \kms, and $1\sigma=474$ \kms)\ around 0 \kms\ due to the uncertainties of the systemic redshifts and the random distribution of intervening systems respectively. The mauve dash curve shows the combination of the two. \label{fig:AAL_v}}
\end{figure}

We show the SNR$_{1700}$ distribution of the quasar spectra and the REW distribution of \civ\ absorption lines from the full sample in \Cref{fig:snr_REW_dis}, and the velocity shift distribution of the \civ\ NALs from the subsample in \Cref{fig:AAL_v}. In \Cref{fig:AAL_v} we have not corrected for the detection completeness variation as a function of velocity shift because the velocity dependence is weak for this dataset. The results shown in \Cref{fig:AAL_v} are in good agreement with previous studies that do make those detection completeness corrections (see Fig.~5 in \citet{Nestor08} and Fig.~9 in \citet{Wild08}). From \Cref{fig:AAL_v}, we can tell that most of the measured NALs are within -1000 to 1000 \kms\ to the quasar redshift, and the velocity shift distribution includes contributions from four different types of NALs, namely, 1) unrelated intervening systems with no significant velocity shift dependence (represented on average by dotted blue line in \Cref{fig:AAL_v}), 2) low-speed NALs that have a nearly Gaussian distribution of measured line shifts centered at v$\,\sim 0$ \kms, where the velocity spread is caused at least in part by uncertainties of the quasar systemic redshifts (see solid red curve in \Cref{fig:AAL_v}), 3) an excess above the Gaussian curve at the positive velocities that could identify gaseous infall, and 4) an excess above both the Gaussian curve and the flat intervening distribution at the negative velocities caused by high-speed outflows ejected from the quasars. It is evident from \Cref{fig:AAL_v} and previous studies \citep[][]{Weymann79,Nestor08,Wild08,Perrotta18} that the majority of strong NALs in the velocity range of $-8000<\,$v$\,<-1000$ \kms\ are formed in the high-speed quasar-driven outflows.

The crux of our study is a statistical comparison of different groups of \civ\ NALs. \Cref{tab:definitions} summarizes the main NAL groups we consider, sorted mainly by velocity shifts, where negative velocities indicate blueshifts relative to the quasar emission-line redshifts and positive velocities indicate redshifts. `AALs' are any NALs from our subsample within velocity shift of -8000 \kms\ to 5000 \kms. We adopt this velocity range because NALs in this range have a high probability of being intrinsic to the quasar environments \citep[\Cref{fig:AAL_v}; see also][]{Weymann79, Nestor08, Wild08}. We define candidate `outflow AALs' as in the more limited velocity range of -8000 \kms\ to -1000 \kms\ again based on \Cref{fig:AAL_v}. Previous studies have shown that most NALs in this velocity range form in quasar-driven outflows. In particular, velocity shifts v$\, \lesssim -1000$ \kms\ are too large to include starburst-driven winds and other ambient gas in the quasar host galaxies \citep{Richards99, Heckman00, Nestor08}. We define candidate `infall AALs' as another subset of AALs at velocity shifts $>400$ \kms. This group aims to isolate true infall systems from those formed in ambient gas that can have a range of measured velocities near v$\,\sim 0$ due to the redshift uncertainties. In particular, v$\,>400$ \kms\ can exclude most of the Gaussian distribution ($1\sigma$) shown in red in \Cref{fig:AAL_v}. 

\begin{deluxetable}{cccc}
\tablecaption{Groups of \civ\ NALs. Quantities listed are the group name, velocity shift range (\kms), number of quasars, and number of NALs in each group.\label{tab:definitions}}
\tablehead{
\colhead{Line Group} & \colhead{Velocity shift} & \colhead{\# Quasars} & \colhead{\# NALs}\\
\colhead{}  & \colhead{(\kms)} & \colhead{} & \colhead{}
}  
\startdata
AALs & $-$8000 to 5000 & 22,335 & 25,264 \\
Outflow AALs & $-$8000 to $-$1000 & 11,097 & 12,213 \\
Infall AALs & $\geq400$ & 2896 & 2945 \\
\enddata
\end{deluxetable}



\begin{figure}
\centering
\includegraphics[width=0.5\textwidth]{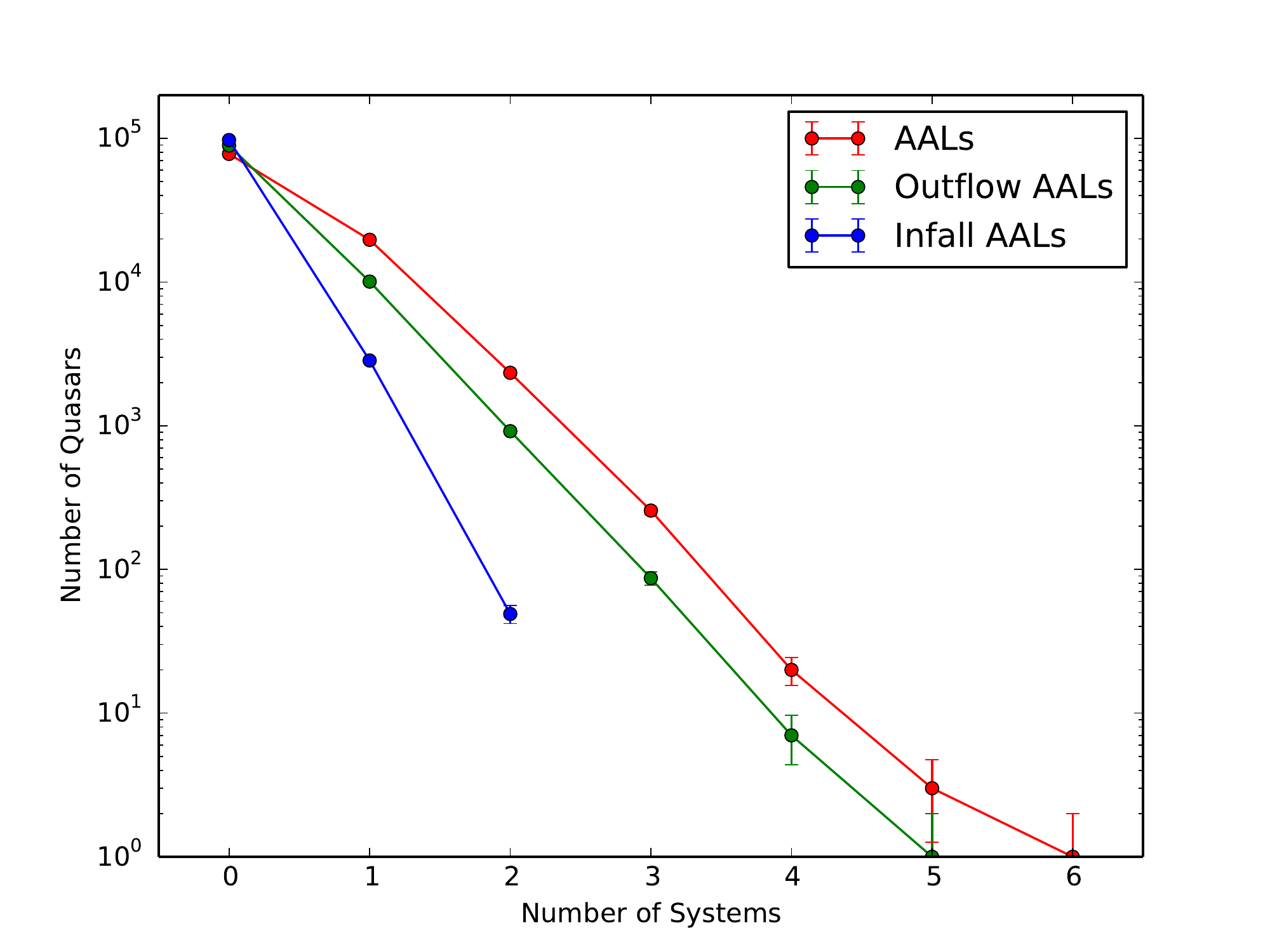}
\caption{The red, green, and blue lines show the quasar number distribution in the number of AALs, outflow AALs, and infall AALs, respectively, contained in each quasar. The error bars indicate the $1\sigma$ uncertainty from counting statistics.\label{fig:number}}
\end{figure}

\Cref{fig:number} shows the quasar number distribution in the number of \civ\ NALs systems contained in each quasar, for AALs, outflow AALs, and infall AALs, respectively. As shown in \Cref{fig:number}, most of them do not contain any of the three NALs groups defined above. It is important to keep in mind that these numbers are limited to relatively strong \civ\ NALs with REW~$>0.5$ \AA\ due to the relatively low resolution ($R\sim 2000$ corresponding to $\sim$150 \kms) and low signal-to-noise ratios in the BOSS spectra. 

An important part of our analysis is to compare the incidence of different groups of \civ\ NALs to \civ\ BALs in the same spectra. We identify BAL and non-BAL quasars (BALQSOs and non-BALQSOs) using the `balnicity index' (BI) as reported in the BOSS DR12 quasar catalog \citep{Paris17}. Specifically, we require $\rm{BI}=0$ for non-BALQSOs and $\rm{BI}\geq500$ measured at $\geq3\sigma$ significance for BALQSOs. We also require that the BAL troughs do not have significant absorption at velocities v$\,> -8000$ \kms . This may eliminate many BALs from our sample, but can avoid the spectral overlap with AALs at low velocities, which is essential for some of our analysis in Section~\ref{sec:correlation}. We enforce this velocity cut by requiring \texttt{vmin\_civ\_2000}~$< -8000$ \kms\ as recorded in the BOSS DR12 quasar catalog \citep[where \texttt{vmin\_civ\_2000} is the lowest velocity where the \civ\ BAL trough dips $>$10\% below the continuum,][]{Paris17}. 
Since it is know that some quasar properties, such as the UV luminosity \citep[e.g.,][]{Gibson09b}, are correlated with the BAL velocity, we have also investigated the impact of this velocity cut of BAL troughs in our study.
We have done analysis without this BAL velocity cut and found little changes to our analysis results presented in Section~\ref{sec:correlation}.

\section{Analysis \& Results \label{sec:correlation}}

\subsection{REW and FWHM Distributions}

\begin{figure*}
\centering
\includegraphics[width=0.8\textwidth]{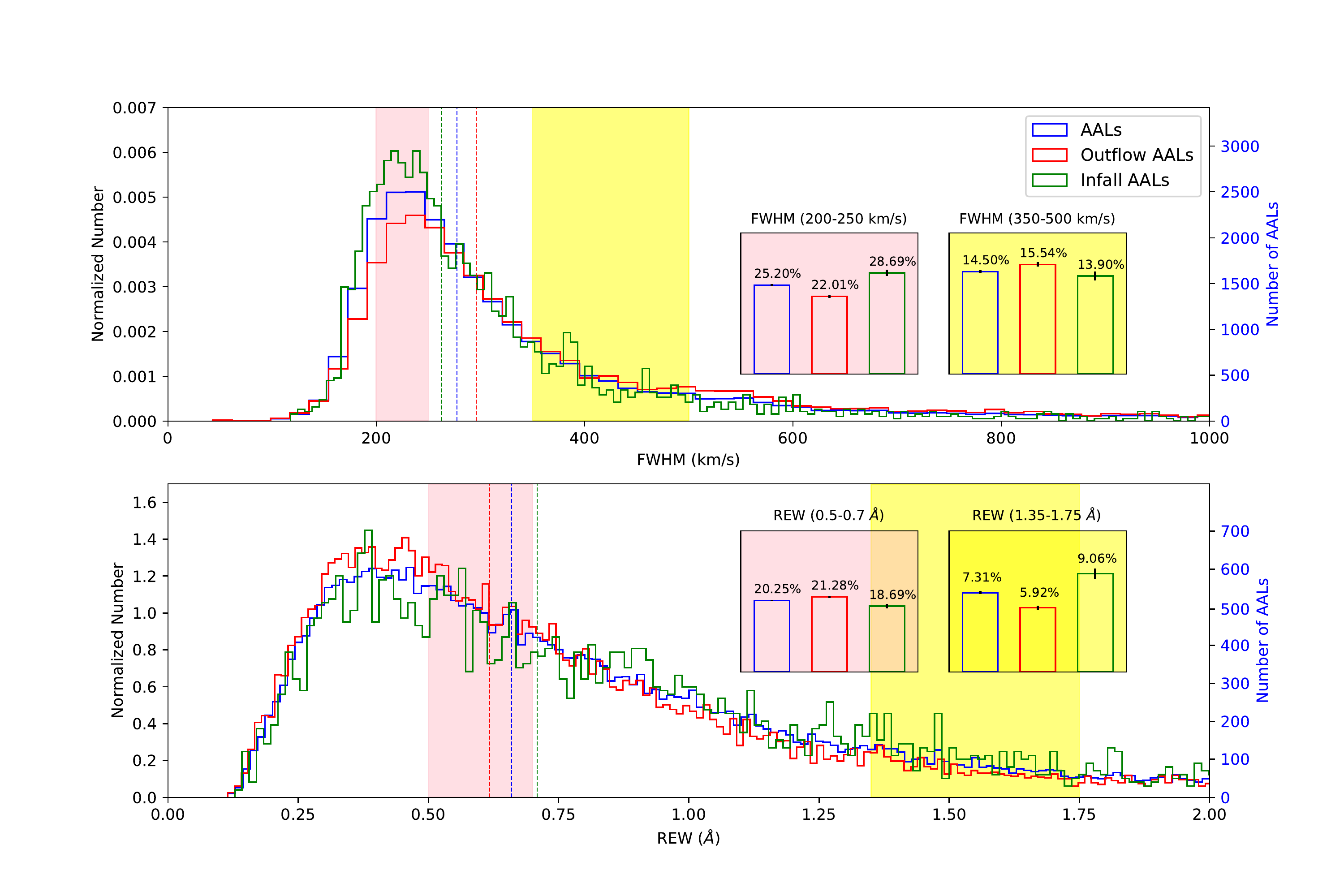}
\caption{
Top panel: Normalized FWHM distributions for AALs, outflow AALs, and infall AALs. The labels of number of AALs in the blue bins are shown on the right axis. Inset: the pink and yellow insets plot the fractionf of NALs within the specific color marked areas in the FWHM distribution. The selection of the color marked areas is described in the text. We do not apply the FWHM $\leq500$ \kms\ cutoffs to the FWHM distributions so as to include the full sample of quasars in these plots. The dashed vertical lines show the median values of FWHM distribution, 262, 277, and 296~\kms, in different NALs groups.
Bottom panel: Normalized REW distributions for AALs, outflow AALs, and infall AALs. The labels of number of AALs in the blue bins are shown on the right axis. Inset: the pink and yellow insets plot the fractions of NALs within the specific color marked areas in the REW distribution. We do not apply the REW $\geq 0.5$ \AA\ cutoffs so as to include the full sample of quasars in these plots. The dashed vertical lines show the median values of REW distribution, 0.62, 0.66, and 0.71~\AA, in different NALs groups.
\label{fig:rew_fwhm}}
\end{figure*}

\Cref{fig:rew_fwhm} compares the normalized FWHM and REW distributions of the \civ\ AALs, outflow AALs, and infall AALs. The distributions are normalized for easier comparisons. 
The median values are 0.66, 0.62, 0.71 \AA\ for the REW distribution of AALs, outflow AALs, and infall AALs respectively, and 277, 296, 262 \kms\ for FWHM distribution of AALs, outflow AALs, and infall AALs respectively.
To study the dependence of the NALs occurrence rate on the FWHM and REW distribution, We also compare the fractions of different NALs groups in the lower end (pink marked area) and upper end (yellow marked area) end of the FWHM and REW distributions respectively. The selection of the lower end (200$-$250 \kms\ for FWHM or 0.5$-$0.7 \AA\ for REW) and upper end (350$-$500 \kms\ for FWHM or 1.35$-$1.75 \AA\ for REW) on the plot is done with the principles of 1) selecting typical area on both ends within the parameter ranges set in section 2, 2) enough number of sources in each area to ensure good statistics, 3) AALs are resolved at the BOSS spectral resolution, and 4) usage of round numbers. The results are shown in the pink and yellow insets of \Cref{fig:rew_fwhm}.
As can be seen from the insets, there is a significant tendency for the infall AALs to be narrower (from the distribution of FWHM) and stronger (from the distribution of REW) than the other AAL groups.

We have also done a Kolmogorov–Smirnov (K-S) test to compare the REW distribution of AALs, outflow AALs, infall AALs in BALQSOs and non-BALQSOs respectively, and found no significant difference between the REW distribution of NALs in BALQSOs and non-BALQSOs. This indicates statistically the presence of BAL components has no significant impact on the strengths (as shown by REWs) of AALs, outflow AALs, or infall AALs in the same quasar.

\subsection{Correlation Analysis}

In the remainder of this section, we use the Z-test to study the correlations between the incidence rates of different \civ\ NAL groups to the intrinsic quasar properties including BALs and radio-loudness. The Z-test for two group fractions is used when determining whether two groups (e.g., BALQSOs and non-BALQSOs) differ significantly on some categorical characteristics (e.g., whether they have AALs) when the variances are known and the sample size is large. For example, when study the correlation between the incidence rate of AALs and the BALs property, we use four numbers as our Z-test inputs, including the number of BALQSOs with and without AALs, and non-BALQSOs with and without AALs. Similar Z-tests have been done in our correlation analysis.  
The results of the Z-tests, including the $Z$ values and P-values, are shown in \Cref{tab:significance}. If $Z\geq 2.58$ or $Z\leq -2.58$ ($P<0.01$), then statistically the incidence rate of specific type of NALs and some intrinsic quasar property (such as BALs or radio loudness) should have strong positive or negative correlation. The P-value here represents the probability that the correlation can occur by a random chance. We will explore these correlation results in detail next. 

\begin{deluxetable}{cccc}
\tablecaption{The Z-test results Z-value (P-values in the parenthesis) of correlation between the incidence rates of different \civ\ NAL groups and intrinsic quasar properties, including the presence of BALs and radio-loudness as described in the text. Value of $\vert Z\vert \geq 2.58$ indicates a strong correlation at $\geq$99\%\ confidence. For example, the Z-value of 3.64 in the first column suggests BALQSOs are more likely to have \civ\ AALs in their spectra than non-BALQSOs. \label{tab:significance}
}
\tablehead{
\colhead{}  & \colhead{AALs} & \colhead{Outflow AALs} & \colhead{Infall AALs}
}  
\startdata
 BALs & 3.64 (0) & 6.84 (0) & -1.47 (0.14)  \\
 Radio loudness & 0.88 (0.38) & 0.37 (0.70) & -0.02 (0.98) \\
\enddata
\end{deluxetable}

\begin{figure*}
\centering
\includegraphics[width=1.0\textwidth]{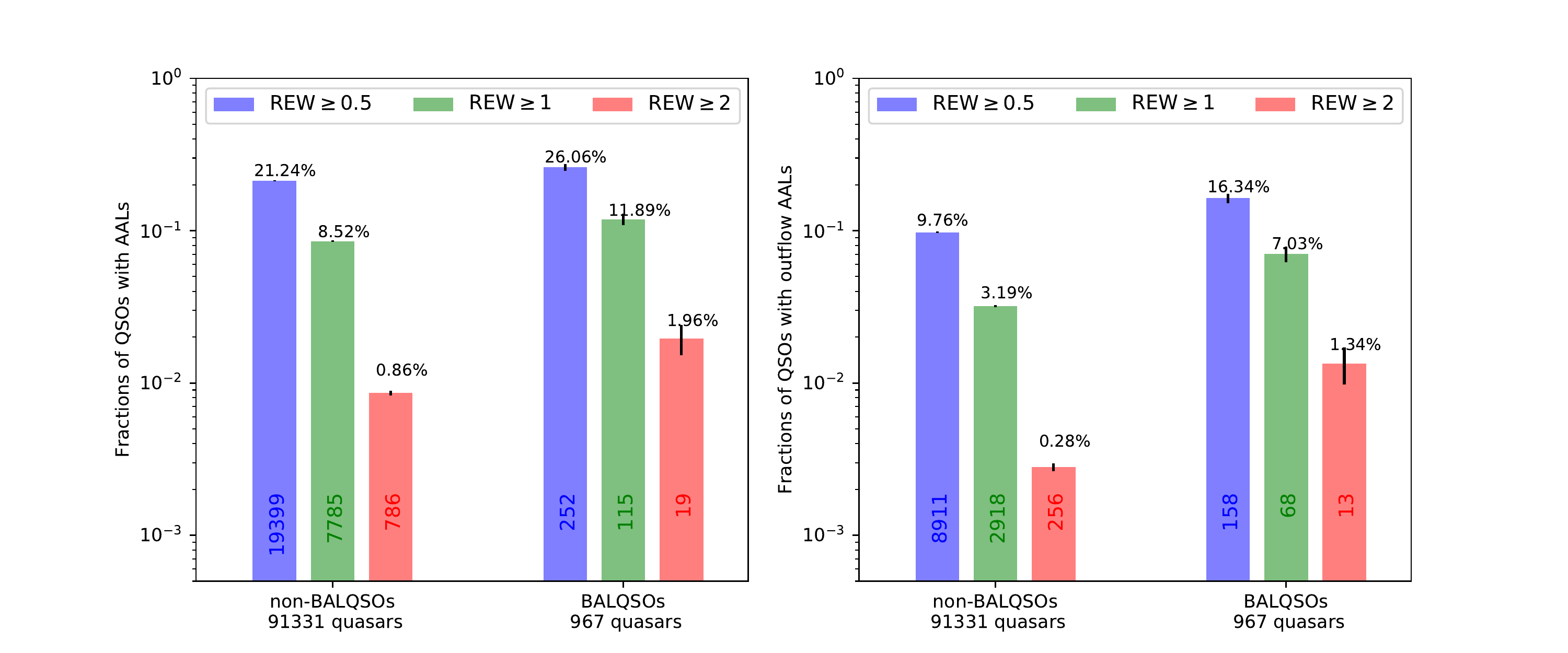}
\caption{Fractions of non-BALQSOs and BALQSOs with AALs (left panel) and outflow AALs (right panel) in different REW thresholds (REW$\geq$0.5, REW$\geq$1, and REW$\geq$2). The number of quasars with AALs or outflow AALs is labeled on each bar, and the fraction of quasars with AALs or outflow AALs is labeled on the top of each bar. The total numbers of non-BALQSOs and BALQSOs included in these plots (91,331 and 967, respectively) are recorded below the panels. The error bars are $1\sigma$ uncertainty from counting statistics.\label{fig:AAL_outflow}}
\end{figure*}

\Cref{fig:AAL_outflow} plots the fractions of quasars with \civ\ AALs and outflow AALs for quasars with and without \civ\ BALs. In this plot, we have separated the \civ\ AALs according to certain REW threshold. From the left panel of \Cref{fig:AAL_outflow}, there are significantly higher probabilities for AALs appearing in BALQSOs compared to non-BALQSOs. From the right panel of \Cref{fig:AAL_outflow}, these differences are larger for outflow AALs. For both AAL groups in \Cref{fig:AAL_outflow}, stronger AALs (larger REWs) show bigger ratios of the fraction percentage. For example, outflow AALs with REW $\geq$ 1 \AA\ are $>$2 times more common in BALQSOs than in non-BALQSOs, and those with REW $\geq$ 2 \AA\ are $>$4 times more common. The results shown graphically in \Cref{fig:AAL_outflow} are supported by the Z-test results in \Cref{tab:significance}. In particular, the Z-test values indicate that both AALs and outflow AALs are strongly correlated with BALs at $>$99.7\% confidence.  

\begin{figure}
\centering
\includegraphics[width=0.5\textwidth]{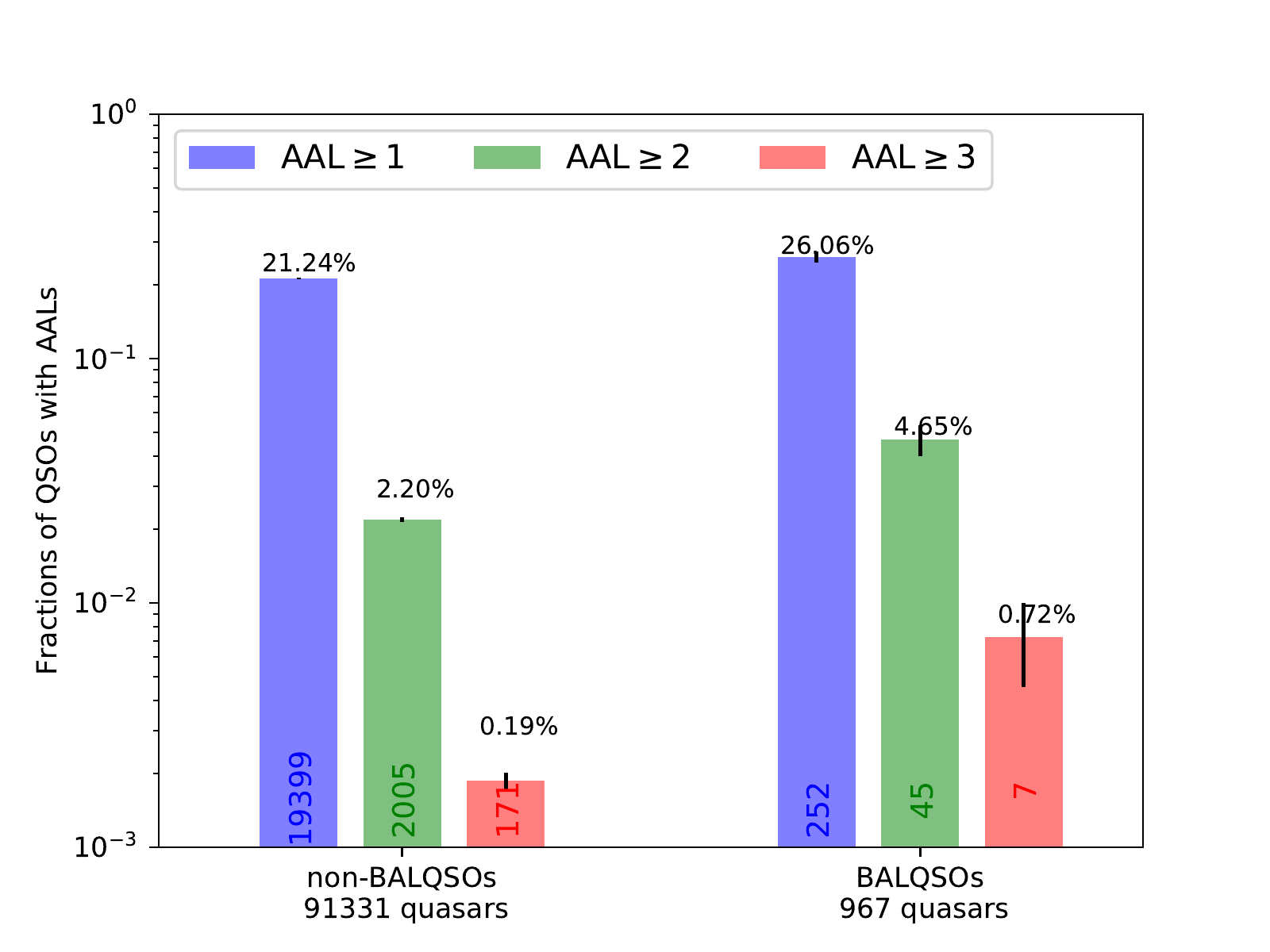}
\caption{Fractions of non-BALQSOs and BALQSOs with different numbers of AALs. The number of quasars with 1, 2 or 3 AALs is labeled on each bar. See \Cref{fig:AAL_outflow} for additional notes.\label{fig:BI_number}}
\end{figure}

\Cref{fig:BI_number} shows the fractions of quasars with different numbers of AALs in BALQSOs versus in non-BALQSOs. Note that the bars showing the fractions with $\geq$1 AAL are the same as bars showing the fraction with REW$\geq$0.5 in the left panel of \Cref{fig:AAL_outflow}. From \Cref{fig:BI_number}, the ratio between the fraction percentage of BALQSOs with AALs and the fraction percentage of non-BALQSOs with AALs is larger regarding multiple AAL systems than regarding single AAL systems. This suggests that BALs play a more important role on the formation of multiple AALs systems.

\begin{figure}
\centering
\includegraphics[width=0.5\textwidth]{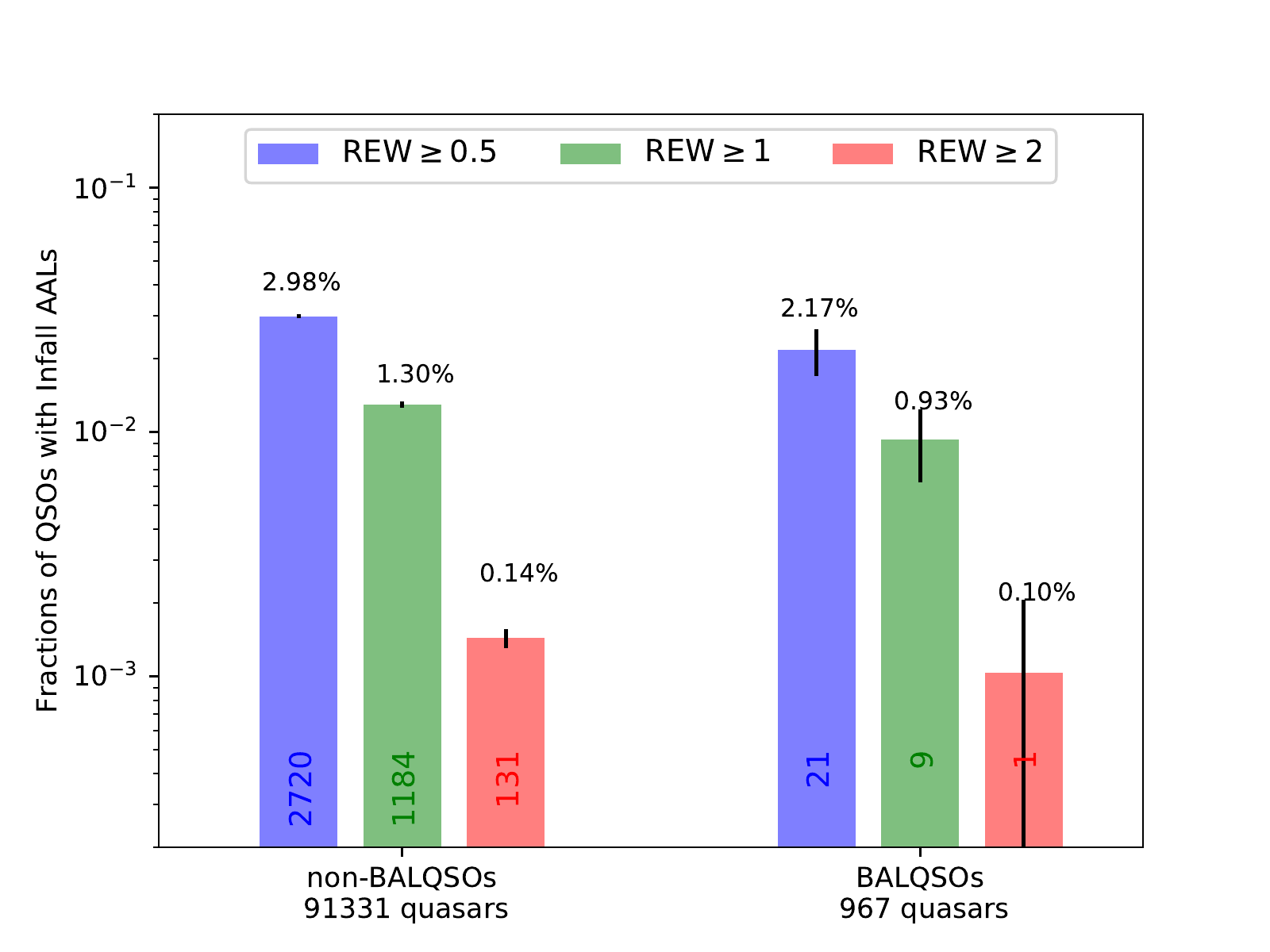}
\caption{Fractions of quasars with infall AALs in both non-BALQSOs and BALQSOs. The number of quasars with infall AALs is labeled on each bar. See \Cref{fig:AAL_outflow} for additional notes.\label{fig:BAL_infall}}
\end{figure}

\Cref{fig:BAL_infall} plots the fractions of quasars with one or more infall AALs in BALQSOs versus non-BALQSOs. There exists a weak tendency for fewer infall AALs appearing in quasars with BALs, which has a Z-value of -1.47. The probability of this trend occurring by random chance is $P=14\%$ from the Z-test (see \Cref{tab:significance}). Since infall AALs are a subset of all AALs, the trend of more AALs existing in BALQSOs shown above in \Cref{fig:AAL_outflow} is actually weakened slightly by the opposite trend shown here for the infall AALs. 

Radio-loudness in quasars is an indicator of relativistic jets, which can produce radio flux via synchrotron emissions. These radio jets might sweep up cooler gas in the quasar host galaxies to produce outflow AALs when viewed along particular line of sight. We compare the fractions of quasars with AALs, outflow AALs, and infall AALs versus radio loudness. We use 20 cm radio fluxes from the FIRST radio survey as recorded in the BOSS DR12 quasar catalog \citep{Paris17}. It has been shown that detection at 20 cm in this survey is equivalent to standard definition of radio loud for quasars at the redshifts and rest-UV magnitudes of our study \citep{Richards01a}. Conversely, non-detection indicates that the quasar is radio-quiet. The Z-test results are listed in \Cref{tab:significance}. We do not find any significant correlation between radio loudness of quasar in our sample to the presence of AALs, outflow AALs, or infall AALs from the same quasar, with the measured probability of correlation occurring by chance $P\gtrsim40\%$.

\begin{deluxetable}{c|cccc}
\tablecaption{REW measurements (units: \AA) of high-ionization and low-ionization NALs in the composite spectra of AALs, outflow AALs, and infall AALs as shown in \Cref{fig:composite_norm}, respectively. \label{tab:rews}
}
\tablehead{
\colhead{}  & \colhead{}  & \colhead{AALs} & \colhead{Outflow AALs} & \colhead{Infall AALs}
}  
\startdata
 \multirow{15}{*}{low ionization}  & \siii\ \lam 1190 & $0.11\pm0.01$ & $0.14\pm0.01$ & $0.09\pm0.02$ \\  
  & \siii\ \lam 1193 & $0.12\pm0.01$ & $0.16\pm0.01$ & $0.11\pm0.01$ \\  
  & \siii* \lam 1194 & $<0.01$ & $<0.01$ & $<0.01$ \\
  & \siii\ \lam 1260 & $0.15\pm0.01$ & $0.16\pm0.01$ & $0.13\pm0.01$ \\  
  & \siii* \lam 1260 & $<0.01$ & $<0.01$ & $<0.01$ \\
  & \oi\ \lam 1302 & $0.05\pm0.01$ & $0.10\pm0.01$ & $0.03\pm0.01$ \\ 
  & \siii\ \lam 1304 & $0.04\pm0.01$ & $0.08\pm0.01$ & $0.03\pm0.01$ \\  
  & \siii* \lam 1309 & $<0.01$ & $<0.01$ & $<0.01$ \\
  & \cii\ \lam 1335 & $0.15\pm0.01$ & $0.18\pm0.01$ & $0.14\pm0.01$ \\  
  & \cii* \lam 1336 & $<0.04$ & $<0.03$ & $<0.06$ \\
  & \siii\ \lam 1527 & $0.09\pm0.01$ & $0.12\pm0.01$ & $0.08\pm0.01$ \\  
  & \siii* \lam 1533 & $<0.01$ & $<0.01$ & $<0.01$ \\
  & \mgii\ \lam 2796 & $0.25\pm0.01$ & $0.40\pm0.02$ & $0.22\pm0.04$ \\ 
  & \mgii\ \lam 2804 & $0.20\pm0.01$ & $0.34\pm0.02$ & $0.24\pm0.04$ \\
 \vspace{2mm} & \mgi\ \lam 2853 & $0.02\pm0.01$ & $0.02\pm0.02$ & $0.05\pm0.04$ \\
\hline
 \multirow{8}{*}{high ionization}  & \ovi\ \lam 1032 & $0.72\pm0.02$ & $0.70\pm0.07$ & $0.78\pm0.04$ \\
 & \ovi\ \lam 1038 & $0.69\pm0.05$ & $0.65\pm0.08$ & $0.74\pm0.06$ \\
 & \nv\ \lam 1239 & $0.43\pm0.01$ & $0.40\pm0.02$ & $0.47\pm0.02$ \\
 & \nv\ \lam 1243 & $0.31\pm0.01$ & $0.27\pm0.01$ & $0.32\pm0.01$ \\
 & \siiv\ \lam 1394 & $0.25\pm0.01$ & $0.25\pm0.01$ & $0.30\pm0.01$ \\
 & \siiv\ \lam 1403 & $0.17\pm0.01$ & $0.17\pm0.01$ & $0.21\pm0.01$ \\ 
 & \civ\ \lam 1548 & $0.90\pm0.01$ & $0.88\pm0.02$ & $0.92\pm0.01$ \\
 & \civ\ \lam 1551 & $0.72\pm0.07$ & $0.71\pm0.08$ & $0.76\pm0.07$ \\  
\enddata
\end{deluxetable}

\subsection{Composite Spectra}

\begin{figure}
\centering
\includegraphics[width=0.45\textwidth]{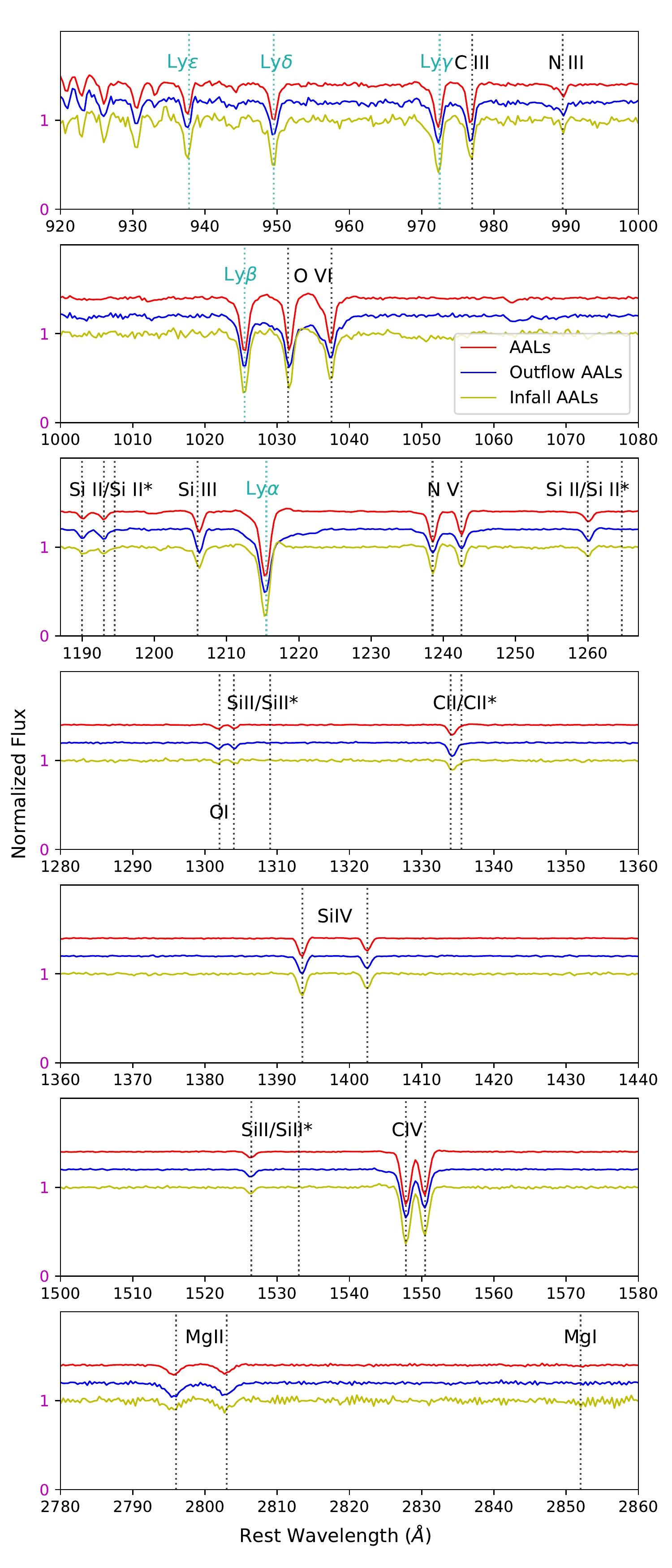}
\caption{Normalized median composite spectra of non-BALQSOs in the absorber frame sorted by NAL groups. The red, blue, and yellow lines show the composite spectra of AALs, outflow AALs, and infall AALs, respectively. Offsets are added to the spectra for displaying purpose.\label{fig:composite_norm}}
\end{figure}

We create median composite spectra in the absorber frame for all non-BALQSOs sorted by NAL group, e.g., AALs, outflow AALs, and infall AALs, respectively. Portions of the resulting normalized composite spectra are shown in \Cref{fig:composite_norm}. We have measured the REWs for the high-ionization and low-ionization lines detected in the composite spectra, and the results are presented in \Cref{tab:rews}. Comparing with the outflow AALs, the infall AALs have moderately stronger high-ionization lines, e.g., \civ, \nv, \ovi, and weaker low-ionization lines, e.g. \siii, \cii. Thus, contrary to our common understanding, this indicates that the infall gas has moderately higher ionizations than quasar-driven outflows. 
There is no significant absorption in any of the excited-state lines like \siii* \lam 1265,1533 and \cii* \lam 1336 from the composite spectra of outflow AALs. We will discuss these results further in Section 4 below.

\section{Discussion}

In Section 3, we have described comparisons between different \civ\ NAL groups and their correlations with certain quasar properties, namely BAL outflows and radio-loudness. We find numerous strong correlations as well as differences in the line ratios (ionizations) that can provide new constraints on the physical nature of quasar outflows and infall absorbers in quasar environments. Here we provide a brief discussion.

\subsection{Outflow AALs}

The velocity distribution of \civ\ NALs in our study (\Cref{fig:AAL_v}) confirms the results from previous work \citep{Weymann79, Nestor08, Wild08} that most AALs in the velocity range $-8000<$~v~$<-1000$ \kms\ form in quasar-driven outflows. Further analysis in Section~3.2 shows that the presence of AALs is strongly correlated with BALs. The correlation is especially strong for outflow AALs and strong AALs with large REWs, which suggests that most AALs form in quasar-driven outflows. We also detect a higher incidence rate of multiple AALs when BALs are present, which indicates that quasar outflows tend to be highly structured, thus can create multiple gas components appearing at different velocities along our line of sight. 

Our composite spectra of outflow AALs show no obvious absorption lines in excited state, like \siii* or \cii*, which indicates that typical outflow AALs have low densities, large distances ($>$few kpc) away from the quasars \citep[see more details in][]{Chen18}.

\subsection{Infall AALs}

Another interesting result of our study is a weak tendency for infall AALs (at measured velocity shifts v$\,>400$ \kms ) to appear less often in quasars with BALs than quasars without BALs. This result provides tentative evidence for BAL outflows acting on large scales in quasar host galaxies where they inhibit the infall of gas from the IGM. It supports theoretical models of galaxy evolution that invoke quasar outflow as an important source of feedback to the host galaxies (Section 1). Our composite spectra show further that the infall AALs have higher ionization than outflow AALs. This result might be surprising if we expect the infall AALs to be physically located farther away from the quasar than outflow AALs. However, if the infall absorbers posess lower gas densities, it could cause higher degree of ionization (larger ionization parameters in photoionization models) in spite of being at larger distances away from the quasar than the outflow AALs. These results need further investigation, which is beyond the scope of the present study. 

\section{Summary}

We use the SDSS-BOSS DR12 database to investigate the nature and origin of infall and outflows in quasar environments by examining the relationships of their \civ\ NALs at positive and negative velocity shifts to other quasar properties such as their BAL outflows, and radio-loudness. Our analysis yields the following results:

1) AALs, outflow AALs, and multiple AALs are strongly correlated with BALs, which suggests most AALs have formed in quasar-driven outflows. Since quasar outflows tend to be highly structured, we often can detect multiple gas components with different velocities along our line of sight (Section 3.2 and 4.1).

2) Our median composite spectra of outflow AALs show no obvious absorption lines in excited states. This indicates that typical quasar-driven outflows have low densities and large distances ($>$few kpc) away from the central quasars (Section 3.3 and 4.1). The outflow AALs may be outer extensions of BAL flows much farther out in the host galaxies, and that these outflows are inhibiting inflow at least along the observed line of sight.

3) Infall AALs appear less often in quasars with BALs than quasars without BALs, which indicates BAL outflows act on large scales in quasar host galaxies and hinder the infall of gas from the IGM. It supports theoretical models of which quasar-driven outflow plays an important role in the feedback to the host galaxies (Section 3.2 and 4.2). 

4) Our median composite spectra show that the infall AALs are more highly ionized than outflow AALs. It could be attributed to the lower densities in the infall absorbers in spite of the larger distances compared to outflow AALs (Section 3.3 and 4.2).

5) Our radio loudness correlation analysis finds none of the NAL types correlate significantly with radio loudness, indicating, in particular, that outflow AALs are not caused by powerful radio jets (Section 3.2).

\section*{acknowledgments}

We thank the anonymous referee for the very useful comments and suggestions. This work was supported by funds from University of California, Riverside, USA; funds from Sun Yat-sen University, China; by grant AST-1009628 from the USA National Science Foundation; by the Fundamental Research Funds for the Central Universities, Sun Yat-sen University. Bo Ma thanks the support of NFSC grant U1931102 and 12073092.

\bibliography{reference}
\bibliographystyle{aasjournal}

\end{document}